\documentclass[12pt]{article}
\usepackage{amsmath}
\usepackage{amssymb}
\usepackage{graphicx}
\usepackage{enumerate}
\usepackage{longtable,booktabs,array}
\usepackage{calc}
\usepackage{needspace}
\usepackage[authoryear,round]{natbib}
\usepackage{url} \usepackage{placeins}
\usepackage{float}
\floatstyle{ruled}
\newfloat{algorithm}{tbp}{loa}
\floatname{algorithm}{Algorithm}
\usepackage[hidelinks]{hyperref}
\usepackage{capt-of}
\makeatletter
\newcommand{\supplementeqref}[1]{\textup{\tagform@{\ref*{#1}}}}
\makeatother
\let\oldsubsection\subsection
\renewcommand{\subsection}{\FloatBarrier\oldsubsection}

\newcommand{\blind}{1}

\begin{document}

\def\spacingset#1{\renewcommand{\baselinestretch}{#1}\small\normalsize} \spacingset{1}

\if1\blind
{
  \renewcommand{\thefootnote}{\fnsymbol{footnote}}\begin{center}
    {\Large\bf Bayesian Learning of Spatially Heterogeneous\\ Causal Structure in Spatial Transcriptomics \par}
    \bigskip
    {\large
    \begin{tabular}[t]{c}
      Kyungwook Back\\  Department of Statistics, Korea University\\
      and\\
      Junsouk Choi\footnotemark[2]\\
      Department of Statistics, Korea University
    \end{tabular}\par}
  \end{center}
  \footnotetext[2]{Corresponding author: \texttt{junsoukchoi@korea.ac.kr}}} \fi

\if0\blind
{
  \bigskip
  \bigskip
  \bigskip
  \begin{center}
    {\Large\bf Bayesian Learning of Spatially Heterogeneous\\ Causal Structure in Spatial Transcriptomics \par}
  \end{center}
  \medskip
} \fi

\bigskip
\begin{abstract}
Spatial transcriptomics preserves the location of each expression profile, making it possible to study how gene regulatory networks differ within a tissue. Across regions of a tissue, regulatory relationships may vary and even reverse direction, so identifying the causal relationships in a specific pathological region can provide insight into therapeutic targets. We propose SpaDAG, a Bayesian method for learning spatially varying causal directed acyclic graphs. By defining the graphs on a triangulation of the tissue, SpaDAG allows each edge to have its own spatial pattern and even to reverse direction across regions. We show that the graphs nonetheless remain fully identifiable, and develop an MCMC sampler that explores them efficiently by reversing an edge over an entire connected region in a single move. SpaDAG outperforms competing methods in simulations with and without direction reversals. In a section spanning a renal cell carcinoma and the adjacent kidney, SpaDAG finds largely different relationships in the two tissues, with those in the tumor concentrated on genes associated with therapeutic targets. Between two of these genes, it identifies a relationship whose direction reverses across the tumor boundary, generating hypotheses about their region-specific coupling. Supplementary materials and code for this article are available online.
\end{abstract}

\vfill

\newpage
\spacingset{1.45}

\section{Introduction}
\label{sec:intro}

\paragraph{Scientific background.} Recovering gene regulatory networks from expression data is a long-standing problem in genomics. Knowing which genes regulate which others helps explain how a cell state is maintained or changed, and where that regulation breaks down in disease. Bulk RNA sequencing measures expression aggregated across cells in a tissue sample, making it difficult to distinguish regulatory relationships specific to different cell types. Single-cell RNA sequencing provides expression profiles for individual cells, allowing regulatory networks to be studied across cell types and states \citep{aibar2017scenic}. However, conventional single-cell RNA sequencing involves tissue dissociation and does not retain the original locations of the cells. This limits the study of how regulatory relationships vary with local tissue context, such as proximity to a tumor boundary. 
Spatial transcriptomics, a recent advance, instead measures expression profiles in an intact tissue section and retains the location of each profile, which comes from an individual cell or from a spot on a grid covering the section, depending on the platform \citep{rao2021exploring}. It thereby reveals the spatial heterogeneity of gene expression, making it possible to study not only which genes are related but also where in the tissue the relationship holds.
\rm

A tissue section contains regions that differ in cell composition and expression programs, such as the two sides of a tumor boundary, different cortical layers, or a site of immune infiltration. The regulatory relationships among genes can differ between these regions as well. A gene pair may be connected in one region and not in another, and for some pairs which gene is upstream depends on the region. 
In zebrafish embryos, for example, the blood-cell regulators gata1 and pu.1 are such a pair: gata1 represses pu.1 in the region where red blood cells first form, whereas pu.1 represses gata1 in the region where white blood cells first form \citep{monteiro2011gata}. 
\rm
We seek to identify where regulatory relationships are present within a tissue and whether their directions differ across locations.

\paragraph{Motivating case study: the clear cell RCC tumor--kidney interface.}\label{sec:intro-rcc} A tumor boundary is a natural setting in which such differences can be investigated, because tumor cells and the adjacent normal cells of the same organ lie within one section. We study this setting in clear cell renal cell carcinoma (ccRCC), the most common subtype of renal cell carcinoma (RCC), which arises from renal tubular epithelial cells \citep{hsieh2017rcc}. 

Most ccRCC tumors carry genetic or epigenetic alterations of the tumor suppressor VHL \citep{nickerson2008vhl}. 
In cells with adequate oxygen, the VHL protein marks the $\alpha$ subunits of the hypoxia-inducible factors (HIFs) for degradation, which prevents these transcription factors from activating the genes of the low-oxygen response \citep{kaelin2008vhl}. When VHL is inactivated, HIF-2$\alpha$, the $\alpha$ subunit encoded by the gene EPAS1, accumulates, supports tumor growth, and increases the expression of HIF target genes such as VEGFA \citep{raval2005hif}. The KEGG RCC pathway (hsa05211, \citealp{kanehisa2000kegg}) contains the genes involved in this process and in the growth-factor signaling connected to it. Therapies for advanced ccRCC target several parts of this pathway, including VEGF signaling and HIF-2$\alpha$ \citep{choueiri2020hif2vegf,choueiri2024belzutifan}. The pathway, however, describes the relationships among its genes for the disease as a whole, without reference to location within a tissue. We ask how these regulatory relationships differ between a tumor and the adjacent kidney tissue, both in whether a relationship is present and in which gene is upstream.
\rm

\begin{figure}[!htbp]
\begingroup
\centering
\includegraphics[width=\linewidth]{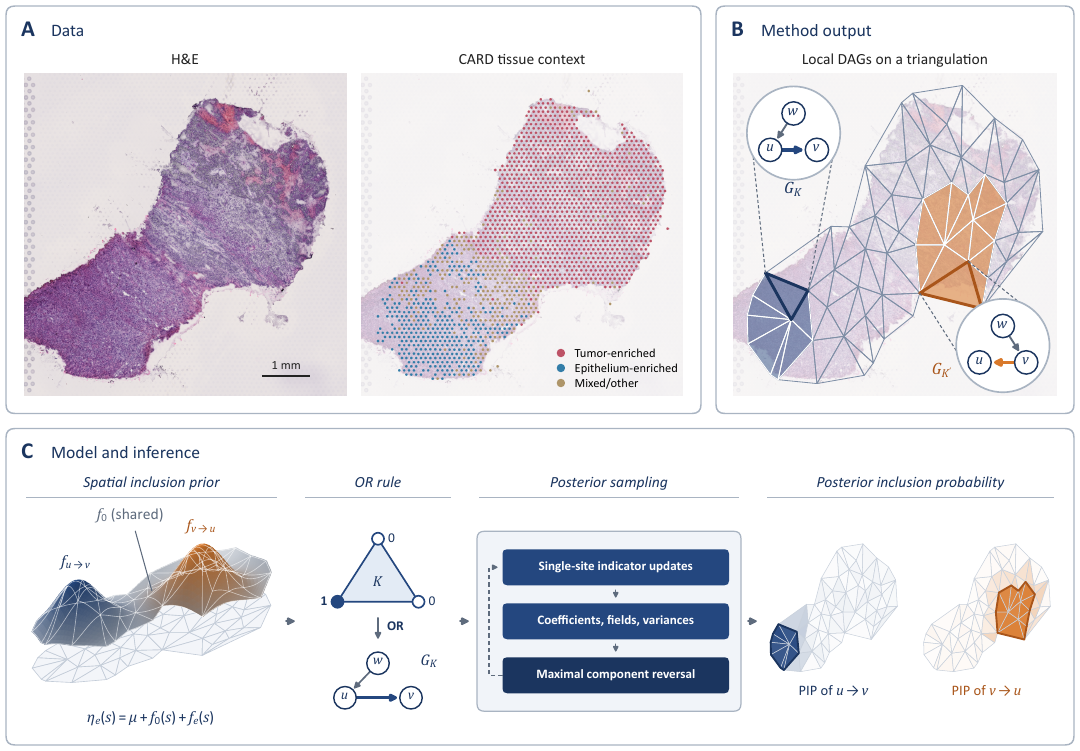}
\caption{Overview of SpaDAG. (A) The ccRCC tumor--kidney interface section: H\&E image and CARD tissue context of the 1,885 retained spots. Scale bar, 1 mm. (B) Local DAGs on a triangulation of the section; shaded triangles carry $u\rightarrow v$ (blue) or $v\rightarrow u$ (orange). (C) The inclusion propensity of each direction combines a shared field $f_0$ with a direction-specific field, vertex indicators give the local DAGs through the OR rule, and the posterior inclusion probability (PIP) of each direction averages the sampled local DAGs. Panels B and C are illustrative.}
\label{fig:overview}
\endgroup
\end{figure}

We analyze a tissue section spanning ccRCC and adjacent kidney tissue, generated by \citet{li2022kidney}, using the counts distributed with the CELINA analysis \citep{shang2025celina}. The section was profiled with the 10x Visium platform, which measures gene expression at spots laid over the tissue, each typically covering several cells. 
We label each spot as tumor-enriched, epithelium-enriched, or mixed/other according to its cell-type proportions, estimated by the deconvolution method CARD \citep{ma2022card}; Figure~\ref{fig:overview}A shows these labels beside the hematoxylin and eosin (H\&E) image. The labels are not an input to our method, which uses only the expression profiles and the coordinates of the spots; they serve to interpret the relationships it estimates.
\rm

\paragraph{Spatially heterogeneous causal structure and existing methods.}\label{sec:intro-methods} 
Which gene is upstream of another, and what would change if a given gene were perturbed, are causal questions. We address them within the framework of causal graphical models, in which the regulatory structure at each location is represented as a directed acyclic graph (DAG) with a causal interpretation. This representation rests on the causal Markov assumption \citep{spirtes2000causation}, which states that each gene is conditionally independent of its non-effects (non-descendants) given its direct causes (parents). The causal interpretation comes from treating the conditional distribution of each gene given its parents as a mechanism that stays the same when other genes are intervened on \citep{pearl2009causality}. Estimating such a graph from observational data is known as causal structure learning, which has served as a tool for generating causal hypotheses in genomics \citep{maathuis2010predicting}.
\rm

With a DAG at each location, the regulatory structure can differ across the tissue in two ways: an edge may be present in one region and absent in another, or it may point in opposite directions in two regions. However, some methods for jointly estimating multiple DAGs allow only the first, because they impose a shared topological ordering, an ordering of the variables in which every parent precedes its children \citep{ni2019bgr,wang2020joint}. Under such an ordering, an edge points the same way wherever it is present. The regional changes in direction that we aim to investigate can therefore be represented only if no single ordering is imposed on the whole tissue.

In addition, nearby locations in a tissue tend to have similar cell composition and expression programs, so their regulatory relationships are likely to be similar as well. Estimating a DAG at each location should rely on this similarity, because a small neighborhood contains few observations relative to the number of candidate edges. 
\rm
Different regulatory relationships, however, need not vary across the tissue in the same way. One relationship may be present throughout the section, while another is confined to the tumor region or changes direction elsewhere. We therefore seek to share information across nearby locations while allowing each edge to have its own spatial pattern.

Existing methods meet some of these requirements. Network methods for spatial transcriptomics use spatial information to estimate graphs that vary over the tissue, but they return undirected graphs \citep{acharyya2025sgr} or regulatory networks that are not causal \citep{li2026svgrn}. Other methods allow the DAG to vary across units or contexts and do not impose a common ordering. Some estimate a DAG for each unit, such as a subject \citep{oates2016multiple}, whereas ContextualDAG \citep{thompson2024contextual} maps context variables to DAGs through a shared neural network. Applied to spatial data, ContextualDAG takes the coordinates as context variables and relates the graphs at nearby locations only implicitly, through the neural network, 
and provides no identifiability guarantee for a causal interpretation. None of these methods combines a causal DAG at each location with an explicit model of spatial dependence.
\rm

Even when information is shared across locations, the data are limited and many graphs explain them almost equally well, so what matters is how strongly the data support each edge and each direction. Bayesian structure learning addresses this by averaging over graphs under their posterior distribution instead of selecting one \citep{madigan1995bayesian}. Bayesian methods have been developed for the graphs of several groups, estimated up to Markov equivalence \citep{castelletti2020multiple}, for DAGs that vary with covariates \citep{ni2019bgr}, and for DAGs indexed by a latent pseudotime, along which an edge can appear, disappear, or change direction \citep{zhou2023bnlte}. In spatial transcriptomics, BNP-DCGx \citep{dawnni2025bnpdcgx} estimates one graph, which may contain directed cycles, for each cluster of spots. It partitions the spots under a prior that depends on their coordinates and lets all edges share the partition, so given the partition every relationship can change only between the same clusters.
These methods quantify the uncertainty about the graph, but none of them does so for a causal DAG at each location with an explicit model of spatial dependence for each edge.
\rm

\paragraph{Our contribution.} We propose SpaDAG (Spatially varying Directed Acyclic Graphs), a Bayesian method that infers a causal DAG at each location of a tissue section from spatial transcriptomics data, without imposing a shared topological ordering. SpaDAG places a local DAG on each triangle of a triangulation of the tissue and requires acyclicity within each triangle only, so the direction of a gene pair can differ between triangles, as in Figure~\ref{fig:overview}B. As outlined in Figure~\ref{fig:overview}C, a spatial prior specific to each edge shares information across nearby locations while letting each edge have its own spatial pattern, and a sampler that can reverse an edge over a connected region in one move yields posterior probabilities for the presence and the direction of each local edge. When the error variance of each gene is constant over space, the local DAGs are identifiable with prior probability one. In simulations, SpaDAG recovered local directed graphs more accurately than the competing methods. In the ccRCC section, tumor-enriched and epithelium-enriched tissue support almost entirely different directed relationships. The estimated direction between EPAS1 and the growth-factor gene HGF differs between a tumor-enriched region and a region outside it, which a model with a shared ordering cannot represent.

Section~\ref{sec:methods} presents the model and the identifiability result, Section~\ref{sec:inference} the posterior inference, Section~\ref{sec:simulation} the simulations, Section~\ref{sec:rcc} the ccRCC analysis, and Section~\ref{sec:discussion} the limitations and extensions. 
Proofs, implementation details, and additional results are given in the Supplementary Material, where all section, equation, figure, and table numbers carry the prefix S.  
\rm
Code implementing SpaDAG is available at \url{https://github.com/KyungWookBack/SpaDAG}.

\section{Bayesian learning of spatially varying causal DAGs}
\label{sec:methods}

SpaDAG is a Bayesian structural equation model whose causal DAG varies over space. Section 2.1 defines the model on a triangulation of the tissue domain, with a local DAG on each triangle. Section 2.2 shows that, with error variances constant over space and spatially varying structural coefficients, these DAGs are fully identifiable from observational data. Section 2.3 specifies a prior that explicitly models spatial dependence among the local DAGs.
\rm

\subsection{The SpaDAG model}
\label{subsec:spatial-sem-fe}
\label{subsec:data-spatial-representation}
\label{subsec:spatial-regression}

We observe the expression of $p$ genes at $n$ spatial locations $s_1,\ldots,s_n$ in a bounded two-dimensional domain $\Omega\subset\mathbb R^2$. Let $X_i=(X_{i1},\ldots,X_{ip})^\top$ denote the expression vector at location $s_i$, and write $\mathcal D=\{(s_i,X_i):i=1,\ldots,n\}$ for the data. To allow regulatory structure to change across space, we introduce a \emph{spatially varying DAG model}: at each location $s\in\Omega$, the causal regulatory relationships among the genes are represented by a directed acyclic graph (DAG) $G(s)=(\mathcal V,E(s))$. The node set $\mathcal V=\{1,\ldots,p\}$ indexes the genes, and the edge set $E(s)\subseteq\mathcal V\times\mathcal V$, which may vary with location $s$, consists of directed edges that form no directed cycle. We denote by $\mathrm{pa}_{G(s)}(j)$ the set of parents of node $j$ in $G(s)$, and by $\mathrm{ch}_{G(s)}(k)$ the set of children of node $k$. The model assumes that, given the location $s_i$, the joint distribution of $X_i$ factorizes according to the causal DAG $G(s_i)$ \citep{lauritzen1996graphical,pearl2009causality}:
\begin{equation}
\label{eq:dag-factorization}
p(X_i \mid s_i)=\prod_{j=1}^p p\bigl(X_{ij} \mid X_{i,\mathrm{pa}_{G(s_i)}(j)},\, s_i\bigr),
\end{equation}
where $X_{i,\mathrm{pa}_{G(s_i)}(j)}=\{X_{ik} : k\in \mathrm{pa}_{G(s_i)}(j)\}$ collects the expression of the parents of gene $j$.

Since $G(s)$ represents the causal structure at location $s$, the factorization \eqref{eq:dag-factorization} is equivalent to a location-wise causal Markov assumption: conditional on $s_i$ and given its parents (its direct regulators) in $G(s_i)$, each gene is independent of its non-descendants, that is, of the other genes that it regulates neither directly nor indirectly. This is a statement about the distribution given location. The distribution of expression pooled over $\Omega$ mixes over locations and need not be Markov with respect to any of the graphs $G(s)$.
As is standard, the causal Markov assumption rests on causal sufficiency \citep{spirtes2000causation}, which we also require location by location: conditional on $s_i$, the $p$ genes have no unmeasured common causes. This does not exclude unmeasured factors that are determined by location, because their influence is absorbed into the dependence of the conditional distributions in \eqref{eq:dag-factorization} on $s_i$. We assume location-wise causal sufficiency throughout the paper.
\rm

The conditional distribution of each $X_{ij}$ in \eqref{eq:dag-factorization} is specified by a spatial structural equation model (SEM) that relates each gene to the remaining genes through spatially varying structural coefficients and a spatial baseline:
\begin{equation}
\label{eq:spadag-sem}
X_{ij}
=
\alpha_j(s_i)
+
\sum_{k\ne j}\beta_{jk}(s_i)X_{ik}
+
\epsilon_{ij},
\qquad
\epsilon_{ij}\overset{\mathrm{ind}}{\sim}N(0,\sigma_j^2).
\end{equation}
Here, the structural coefficient $\beta_{jk}(s)$ corresponds to the direction $k\to j$ and quantifies the direct causal effect of gene $k$ on gene $j$ at location $s$. The baseline $\alpha_j(s)$ accommodates spatial shifts in mean expression, including those associated with tissue composition and expression programs. The error variance $\sigma_j^2$ is gene-specific and constant over the domain $\Omega$.
Under the spatial SEM, the spatially varying DAG $G(s)$ is determined by which structural coefficients are nonzero at $s$: we assume that $\beta_{jk}(s)\neq 0$ if and only if $k\to j \in E(s)$.
\rm

We represent the coefficient surfaces and baselines by spatial regression splines over a triangulation \citep{sangalli2013ssr}. Let $\mathcal T$ be a triangulation of $\Omega$ with nodes $\xi_1,\ldots,\xi_H$, and let $\{\psi_h\}_{h=1}^H$ be a nodal basis of continuous, piecewise-polynomial functions on $\mathcal T$, with $\psi_h(\xi_g)=\mathbf 1\{h=g\}$.
For a piecewise-linear basis, the nodes are the vertices of $\mathcal T$ and $\psi_h$ is the hat function that equals one at $\xi_h$ and zero at every other vertex; for a piecewise-quadratic basis, the nodes also include the midpoints of the triangle edges, so that each triangle carries six nodes. We then set
\rm
\begin{equation}
\label{eq:spadag-fe-surfaces}
\beta_{jk}(s)=\sum_{h=1}^H b_{jkh}\psi_h(s),
\qquad
\alpha_j(s)=a_{j0}+\sum_{h=1}^H a_{jh}\psi_h(s).
\end{equation}
Together, \eqref{eq:dag-factorization}--\eqref{eq:spadag-fe-surfaces} define the SpaDAG model, in which learning the spatially heterogeneous causal graph $G(s)$ reduces to inference on the finite set of basis coefficients $b_{jkh}$.
\rm

The nodal property makes $b_{jkh}$ the value of $\beta_{jk}$ at node $\xi_h$. On each triangle, a surface is determined by its values at the nodes of that triangle, and adjacent triangles share the nodes on their common side, which makes the surfaces continuous. A surface is therefore identically zero on a triangle if and only if its values at the nodes of that triangle are all zero. This property underlies the local graph representation of Section~\ref{subsec:local-dag-construction}, which recasts DAG learning as a coefficient selection problem. \rm To keep the exposition simple, we work with piecewise-linear basis functions throughout the paper. The nodes are then the vertices of $\mathcal T$ and each surface is affine on every triangle. Higher-degree nodal bases can be used in the same way.

The triangulation is constructed from spatial information alone and is fixed before posterior sampling; the constructions used in this paper are described in Section~\ref*{subsec:supp-triangulation}. The triangulation also determines the spatial resolution of the local graphs, whose sensitivity to mesh construction is evaluated in Section~\ref{sec:simulation}.

\subsection{Identifiability}
\label{sec:identifiability}
\label{subsec:identifiability}

A causal DAG model is identifiable if the distribution of the observed variables determines its graph uniquely. In general it does not: DAGs that imply the same conditional independence relations form a Markov equivalence class and, without restrictions on the structural equations or the error distributions, generate the same set of distributions \citep{verma1990equivalence}. The members of a class connect the same pairs of variables but can orient some of the edges differently, so observational data leave those directions undetermined.
\rm
Identifying the directions of the edges therefore requires
some asymmetry in the model. In nonlinear additive noise models, the
forward and backward fits are generically distinguishable
\citep{hoyer2009anm}; in linear models, non-Gaussian errors provide the
asymmetry \citep{shimizu2006lingam}.

To illustrate the asymmetry we use, we first consider an SEM defined on a single, fixed DAG whose coefficients are allowed to vary. Let
$G$ be a DAG on $\mathcal V$ and $\mathcal U$ an arbitrary set. At each
$u\in\mathcal U$,
\begin{equation}
\label{eq:ident-varying-sem}
X_j=\alpha_j(u)+\sum_{k\in\mathrm{pa}_G(j)}\beta_{jk}(u)X_k+\epsilon_j,
\qquad
\epsilon_j\overset{\mathrm{ind}}{\sim}N(0,\sigma_j^2),
\qquad j=1,\ldots,p,
\end{equation}
where the intercepts $\alpha_j$ and the coefficients $\beta_{jk}$ are
real functions on $\mathcal U$, and $\beta_{jk}$ is not identically zero
on $\mathcal U$ for any edge $k\to j$ of $G$. The graph is thus the same
at every $u$, while the coefficients along its edges may change with $u$
and the error variances do not.

\medskip
\noindent\textbf{Condition 1.}\ \textit{For every $\ell\in\mathcal V$
and every non-empty subset $S\subseteq\mathrm{ch}_G(\ell)$, the sum
$\sum_{j\in S}\beta_{j\ell}(u)^2/\sigma_j^2$ is not constant in
$u\in\mathcal U$.}

\medskip
\noindent\textbf{Proposition 1.}\ \textit{Under Condition~1, the DAG $G$ and
the intercepts, coefficients, and error variances
of~\eqref{eq:ident-varying-sem} are identifiable from the distributions
of $X$ at $u\in\mathcal U$.}

\medskip
In BN-LTE \citep{zhou2023bnlte},
$u$ is a latent pseudotime along which the DAG changes, and the model is
of the form~\eqref{eq:ident-varying-sem} on each set of pseudotimes that
share a DAG. Assuming faithfulness in addition, BN-LTE identifies its
DAGs by identifying each DAG on its own set of pseudotimes, which
requires Condition~1 for every DAG.

In SpaDAG, the graph $G(s)$ is the same at every location of a triangle
$K\in\mathcal T$; we write $G_K$ for this graph.
Model~\eqref{eq:spadag-sem} is then of the
form~\eqref{eq:ident-varying-sem} on each triangle, with $\mathcal U=K$
and $G=G_K$, and the error variances are shared by all triangles.

\medskip
\noindent\textbf{Theorem 1.}\ \textit{If Condition~1 holds on some
triangle $K_0$, then the DAGs $G_K$ for all $K\in\mathcal T$, and the
intercepts, coefficients, and error variances of
model~\eqref{eq:spadag-sem} are identifiable from the conditional
distributions of $X$ given $s$, $s\in\Omega$.}

\medskip
Condition~1 on $K_0$ determines every error variance. Since the
variances are shared, they act as known variances on every other
triangle, which makes the graphs and coefficients identifiable.
Theorem~1 thus requires Condition~1 on a single triangle rather than for
every graph. The proofs of Proposition~1 and Theorem~1 are given in
Section~\ref*{sec:supp-identifiability}.

The result rests on the assumption that each error variance is constant
over space. It extends to variances that change by a factor common to
all genes,
$\sigma_j^2(s)=c(s)\sigma_j^2$
(Section~\ref*{subsec:supp-ident-remarks}), but not to changes in noise
level that differ between genes.

\subsection{Prior specification}
\label{subsec:prior-modeling}

The prior expresses our assumption that local graphs are sparse and tend
to be similar at nearby locations, while an individual edge may still
change or reverse direction from one region to another.

\subsubsection{Local DAGs via a spike-and-slab prior}
\label{subsec:local-dag-construction}
\label{subsec:triangle-dags}

SpaDAG defines local edge presence according to whether the corresponding coefficient is identically zero on a triangle. We encode the structural activity of each nodal coefficient with a binary indicator.

Let $\mathcal E_c=\{(j,k):j\ne k\}$ denote the candidate directions, where $e=(j,k)$ represents $k\to j$ and $\bar e=(k,j)$ its reverse. At vertex $\xi_h$, the indicator $\gamma_{eh}\in\{0,1\}$ records whether the corresponding coefficient $b_{eh}=b_{jkh}$ is structurally active. A point-mass spike-and-slab prior sets inactive coefficients to zero and assigns a Gaussian slab to active coefficients:
\begin{equation}
\label{eq:spadag-population-spike-slab}
b_{eh}\mid\gamma_{eh},\nu_1
\sim
(1-\gamma_{eh})\delta_0
+
\gamma_{eh}N(0,\nu_1).
\end{equation}
Here $\delta_0$ is a point mass at zero and $\nu_1$ is a slab variance
shared by all directions and vertices.

Since a nodal coefficient drawn from the slab equals zero with
probability zero, a coefficient surface is identically zero on a
triangle if and only if its three
nodal indicators are all zero, almost surely. SpaDAG therefore defines
triangle-level edge presence through the OR rule applied to these
indicators. With $V(K)$ denoting the vertex indices of triangle $K$,
\begin{equation}
\label{eq:triangle-graph}
\begin{aligned}
z_{eK}(\gamma)
&=
\max_{h\in V(K)}\gamma_{eh},\\
E_K(\gamma)
&=
\{k\to j:z_{(j,k)K}(\gamma)=1\},
\qquad
G_K(\gamma)=(\mathcal V,E_K(\gamma)).
\end{aligned}
\end{equation}
The graph is thus constant within each triangle and piecewise constant
over the domain, whereas the coefficient surface is continuous and
piecewise linear.

To guarantee a DAG at each location, we restrict the indicators to
configurations for which every induced triangle graph is acyclic:
\begin{equation}
\label{eq:spadag-admissible}
\mathcal A
=
\left\{
\gamma\in\{0,1\}^{|\mathcal E_c|\times H}:
G_K(\gamma)\text{ is a DAG for every }K\in\mathcal T
\right\}.
\end{equation}
We set $G(s)=G_K(\gamma)$ for $s\in K$ off the triangle boundaries,
a measure-zero set.
The constraint applies separately to each triangle and requires no common
topological ordering across the domain. The SEM then defines a joint
Gaussian distribution at every location, and its means and covariances
vary continuously across space
(Section~\ref*{subsec:supp-working-posterior}).

\begin{figure}[!htbp]
\centering
\includegraphics[width=0.87\linewidth]{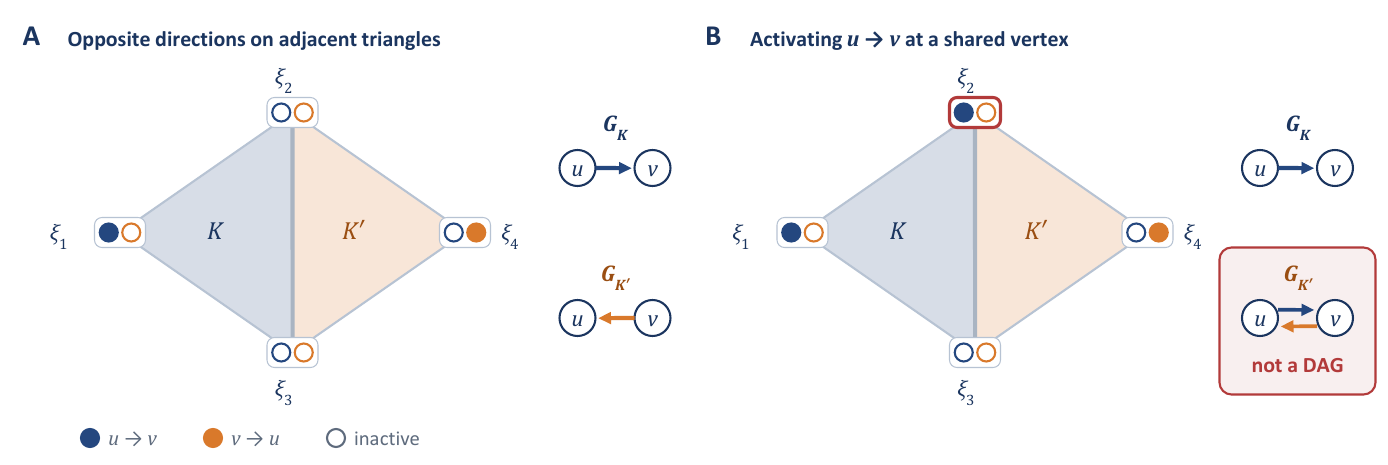}
\caption{Opposite directions of $(u,v)$ on adjacent triangles $K$ and $K'$. (A) With both directions inactive at the shared vertices, $G_K$ contains $u\to v$ and $G_{K'}$ contains $v\to u$. (B) Activating $u\to v$ at $\xi_2$ adds it to $G_{K'}$ through the OR rule, creating a two-cycle.}
\label{fig:orientation-change}
\end{figure}

Local acyclicity and coefficient continuity also constrain how a
direction changes across space. For a continuous coefficient field with
an acyclic graph at every location, any path connecting opposite
orientations of a pair $(u,v)$ passes through a point at which both
directional coefficients vanish. In SpaDAG, this transition can occur on
the boundary shared by two adjacent triangles. If $K$ and $K'$ carry
$u\to v$ and $v\to u$, respectively, neither direction can be active at
their shared vertices, since a direction active there would belong to
both triangle graphs; under the point-mass spike, both coefficients
then vanish along the shared boundary. Each direction can still be active
at the vertex of its triangle opposite that boundary, so the two
directions can occupy adjacent triangles, with no triangle between them
on which the pair is unconnected.
Figure~\ref{fig:orientation-change} shows this constraint on two adjacent
triangles: in panel A both triangle graphs are acyclic, whereas
activating $u\to v$ at a shared vertex in panel B creates a two-cycle in
$G_{K'}$.

This graph representation couples coefficient values and edge support on
adjacent triangles through shared vertices. We next introduce a spatial
inclusion prior to share information about edge presence over a broader
spatial range.

\subsubsection{Spatial inclusion prior}
\label{subsec:spatial-inclusion-prior}
\label{subsec:common-gp-prior}

The spike-and-slab prior~\eqref{eq:spadag-population-spike-slab} is conditional on the nodal indicators, and we now specify the prior of the indicators themselves. We describe it first without the acyclicity restriction~\eqref{eq:spadag-admissible}, which is imposed at the end of this subsection.
SpaDAG assigns each candidate direction $e \in \mathcal{E}_c$ a spatially varying inclusion propensity $q_e(s) \in (0,1)$, $s \in \Omega$, and each nodal indicator $\gamma_{eh}$ follows a Bernoulli prior with the propensity at its vertex $\xi_h$,
\begin{equation}
\label{eq:spadag-indicator-prior}
\gamma_{eh} \mid q_e \overset{\mathrm{ind}}{\sim} \mathrm{Bernoulli}\{q_e(\xi_h)\}.
\end{equation}
Since the propensity is specific to the direction, each edge can have its own spatial pattern.

The logit of each propensity, $\eta_e(s) = \mathrm{logit}\{q_e(s)\}$, combines an overall inclusion level with two spatial components:
\begin{equation}
\label{eq:spadag-inclusion-logit}
\eta_e(s) = \mu + f_0(s) + f_e(s).
\end{equation}
The parameter $\mu$ controls the overall inclusion level. The field $f_0$ is shared by all candidate directions; since it shifts their propensities together, it marks regions where the graph is relatively dense or sparse. The field $f_e$ is specific to direction $e$ and captures the departure of that direction from this common trend.

We assign the fields zero-mean Gaussian process priors, $f_0 \sim \mathrm{GP}(0, \sigma_0^2 \kappa)$ and $f_e \sim \mathrm{GP}(0, \tau^2 \kappa)$, with centering constraints that separate $\mu$, $f_0$, and the $f_e$ (Section~\ref*{subsec:supp-spatial-prior-details}). The common kernel $\kappa(s, s')$ is a spatial covariance kernel on $\Omega$ under which the values of a field at two locations are more strongly correlated the closer the locations are.  The propensities of a given direction are therefore similar at nearby vertices, and the prior favors configurations in which nearby locations share edges. The field variances $\sigma_0^2$ and $\tau^2$, which are learned from the data, determine how strongly the propensities vary over space. The spatial scale of the variation is fixed by the kernel (Section~\ref*{subsec:supp-spatial-prior-details}). The kernel used in this paper is given in Section~\ref{sec:inference}.

The independent Bernoulli distributions in~\eqref{eq:spadag-indicator-prior} do not enforce acyclicity: they give positive probability to configurations in which a triangle graph contains a directed cycle. Let $\vartheta$ collect the overall inclusion level, the values of the fields at the mesh vertices, and the field variances. We define the joint prior of $(\gamma, \vartheta)$ by restricting the distribution specified above to the indicator configurations in $\mathcal{A}$, the set defined in~\eqref{eq:spadag-admissible}:
\begin{equation}
  \label{eq:spadag-joint-inclusion-prior}
  \pi(\gamma, \vartheta) \propto \pi_{\mathrm{base}}(\vartheta)\, \mathbf{1}\{\gamma \in \mathcal{A}\}
  \prod_{e \in \mathcal{E}_c} \prod_{h=1}^{H} q_e(\xi_h)^{\gamma_{eh}} \{1 - q_e(\xi_h)\}^{1 - \gamma_{eh}},
\end{equation}
where the double product is the probability of $\gamma$ under~\eqref{eq:spadag-indicator-prior} and $\pi_{\mathrm{base}}(\vartheta)$ is the prior of $\vartheta$ before the restriction, given by the Gaussian process priors above together with the priors of $\mu$ and the field variances (Section~\ref*{subsec:supp-spatial-prior-details}). Restricting the joint distribution of $(\gamma, \vartheta)$, rather than the conditional distribution of $\gamma$ given $\vartheta$, avoids a normalizing constant over $\mathcal{A}$ that depends on the fields.

\rm
\subsubsection{Identifiability under the SpaDAG prior}
\label{subsec:prior-identifiability}

Under the SpaDAG prior of
Sections~\ref{subsec:local-dag-construction}--\ref{subsec:spatial-inclusion-prior},
the triangle graph $G_K(\gamma)$ in~\eqref{eq:triangle-graph} is almost
surely the DAG $G_K$ of Section~\ref{subsec:identifiability}.

\medskip
\noindent\textbf{Corollary 1.}\ \textit{Under the SpaDAG prior,
Condition~1 holds on every triangle with probability one, and the local
DAGs, coefficients, and error variances are identifiable.}

\medskip
Under our spatial regression spline representation~\eqref{eq:spadag-fe-surfaces} the coefficients are
polynomials on each triangle, for which Condition~1 holds if and only if
no edge has a coefficient that is constant on the triangle
(Section~\ref*{subsec:supp-ident-remarks}). 
A coefficient is constant on a triangle only if the edge is active at every node of that triangle with all nodal values equal, which has probability zero under the Gaussian slab.
The argument holds
for any prior on the indicators; the proof is given in
Section~\ref*{subsec:supp-ident-corollary}.

\section{Posterior inference}
\label{sec:inference}
\label{subsec:inference-gp}
\label{subsec:posterior-inference}
\label{subsec:posterior-sampling}
\label{subsec:posterior-summaries}

This section describes posterior inference for the proposed SpaDAG model. Since the posterior distribution  is intractable, we sample from it using Markov chain Monte Carlo (MCMC). The main difficulty is that the sampler must explore indicator configurations that are acyclic on every triangle and coupled across neighboring vertices. To explore this constrained space efficiently, we develop a Metropolis-within-Gibbs sampler that updates the indicators through two complementary transitions: single-site Gibbs updates and a blocked Metropolis--Hastings move that reverses an edge over a connected region. The remaining parameters are drawn from their full conditional distributions (Section~\ref*{sec:supp-sampler}).

\rm

\paragraph{Maximal component reversal.}
Single-site updates change the orientation of an edge over a connected
region only slowly. The reverse direction cannot be activated at a
vertex while the current direction is active at any vertex of its
incident triangles (Figure~\ref{fig:orientation-change}B). Reversing an
edge over a region one vertex at a time therefore passes through
configurations in which the edge is absent on part of the region. The difficulty
is that, when the data support a directed relationship throughout the region, these
intermediate configurations can have much lower posterior probability
than configurations supporting either orientation.

The blocked move below circumvents this difficulty: acting on the nodal indicators, it reverses the direction over an entire connected component in a single proposal.
Two mesh vertices are adjacent if they
share a triangle. Starting from a vertex at which direction $e$ is
active, let $C$ be the maximal connected component of $e$-active
vertices containing it. Since $\gamma\in\mathcal A$, $\bar e$ is
inactive at every vertex of $C$, and we propose
\begin{equation}
(\gamma_{eh},\gamma_{\bar e h})=(1,0)
\;\longrightarrow\;
(\gamma'_{eh},\gamma'_{\bar e h})=(0,1),
\qquad h\in C,
\label{eq:pi-component-reversal}
\end{equation}
leaving all other indicators unchanged. By the maximality of $C$, every
$e$-active vertex that shares a triangle with a vertex of $C$ belongs to
$C$, so the reversal cannot create a two-cycle between $e$ and $\bar e$
under the OR rule in Section \ref{subsec:local-dag-construction}. Proposals that create a longer directed cycle in any
triangle are rejected. The acceptance probability integrates out the coefficient vectors of
$e$ and $\bar e$, which are redrawn after either outcome, and the
transition satisfies
detailed balance with respect to our posterior distribution (Proposition~\ref*{thm:supp-reversal}).

\paragraph{Gaussian process approximation.}
For scalable posterior computation under the GP priors, we represent each GP through a low-rank basis expansion derived from the eigendecomposition of its covariance kernel. Mercer's theorem gives $\kappa(s,s')=\sum_{\ell\ge1}\lambda_\ell\phi_\ell(s)\phi_\ell(s')$, with eigenvalues $\lambda_1\ge\lambda_2\ge\cdots>0$ and the associated orthonormal eigenfunctions $\{\phi_\ell\}_{\ell\ge1}$. Accordingly, a mean-zero GP with kernel $\kappa$ admits the Karhunen--Lo\`eve expansion $\sum_{\ell\ge1}\theta_\ell\sqrt{\lambda_\ell}\,\phi_\ell(s)$, where the coefficients $\theta_\ell$ are independent Gaussian. Retaining only the $L$ leading terms, we approximate each field as
\begin{equation}
\label{eq:spadag-kl-fields}
f_0(s)
=
\sum_{\ell=1}^L
\theta_{0\ell}\sqrt{\lambda_\ell}\,\phi_\ell(s),
\qquad
f_e(s)
=
\sum_{\ell=1}^L
\theta_{e\ell}\sqrt{\lambda_\ell}\,\phi_\ell(s),
\end{equation}
with $\theta_{0\ell}$ and $\theta_{e\ell}$ independent Gaussian a priori.  
Inference on each field thus reduces to inference on a finite-dimensional Gaussian coefficient vector. In both the simulations and the real data analysis, we use the modified squared-exponential kernel
\begin{equation}
\label{eq:spadag-kernel}
\kappa(s,s')=\exp\{-a(\|s\|^2+\|s'\|^2)-b\|s-s'\|^2\},
\qquad a,b>0,
\end{equation}
whose Mercer expansion $\kappa(s,s')=\sum_{\ell\ge1}\lambda_\ell\phi_\ell(s)\phi_\ell(s')$ admits closed-form eigenvalues and eigenfunctions \citep{lin2024lasir}. The kernel parameters and the truncation level $L$ are fixed in advance (Section~\ref*{subsec:supp-spatial-prior-details}).
\rm

\paragraph{Posterior summaries.}
Since the local graphs are defined on triangles, we summarize
the posterior at the triangle level. For each direction $e$ and triangle $K$, let $z_{eK}(\gamma^{(m)})=\max_{h\in V(K)}\gamma_{eh}^{(m)}$ denote the OR-rule indicator in the $m$-th retained posterior draw, $m=1,\ldots,N_{\mathrm{ret}}$. The posterior inclusion probability (PIP) is estimated by the average of these indicators,
\begin{equation}
\label{eq:spadag-triangle-pip}
\widehat\rho_{eK}
=
\frac{1}{N_{\mathrm{ret}}}
\sum_{m=1}^{N_{\mathrm{ret}}}
\max_{h\in V(K)}\gamma_{eh}^{(m)}.
\end{equation}
These probabilities form spatial maps of posterior support for
individual directions. Since no sampled triangle graph contains both
directions of a pair, $\widehat\rho_{eK}+\widehat\rho_{\bar eK}\le1$;
their sum is the posterior probability that the pair is connected on
$K$, and their comparison shows how this probability is divided between
the two orientations.

\section{Simulation studies}
\label{sec:simulation}

We first examine whether SpaDAG recovers local graphs whose orientations reverse across the domain. The true nodal indicators are specified on a regular triangulation of $[0,1]^2$ with 50 triangles, and data are generated from the model~\eqref{eq:dag-factorization}--\eqref{eq:spadag-fe-surfaces} at locations drawn uniformly from the domain. We consider graph fields in which some edges point in opposite directions in two patches of the domain, while every local graph remains acyclic, so that no topological ordering is shared across the domain. Coefficients are scaled so that the median population signal-to-noise ratio (equation~\supplementeqref{eq:sim-population-snr}) equals one. Taking $n=500$ and $p=50$ as the baseline, we vary one factor at a time, $p\in\{25,75\}$ and $n\in\{250,1000\}$, with 30 datasets per configuration. SpaDAG is fitted on the generating triangulation with a single MCMC chain of 50,000 iterations initialized at the empty graph, and edges are selected according to the median probability model (i.e., PIP $\ge0.5$). Recovery is evaluated through the true positive rate (TPR), false discovery rate (FDR), and $F_1$ score of the edges selected on each triangle, where an edge counts as correct only if its direction is also correct (equation~\supplementeqref{eq:sim-metrics}). The design and fitting protocols are given in Section~\ref*{sec:supp-simulation}.

We compare SpaDAG with Independent NOTEARS \citep{zheng2018notears} fitted to each triangle separately, Joint NOTEARS, which pools edge selection across triangles through a group penalty, JointGES \citep{wang2020joint}, which estimates the triangle graphs under a common topological ordering, and ContextualDAG \citep{thompson2024contextual}, which takes spatial coordinates as context in a shared neural network. As in its authors' simulations, ContextualDAG is tuned with the true number of edges, and we also report a variant given the true location-specific ordering.

\begin{figure}[!tbp]
\centering
\includegraphics[width=\linewidth]{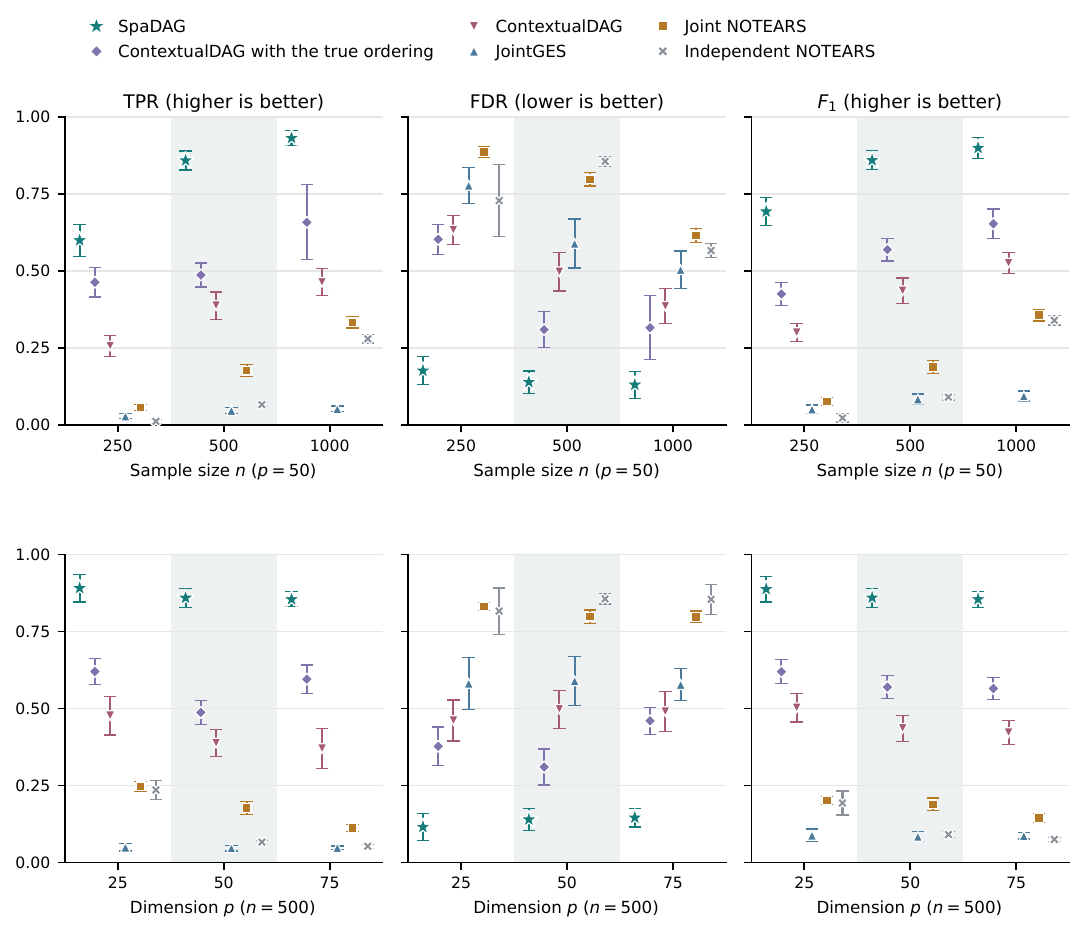}
\caption{Recovery of local directed graphs under spatial orientation reversals: mean TPR, FDR, and $F_1$ score over 30 datasets per configuration, with bars of one between-dataset standard deviation. Top row, $n$ varied at $p=50$; bottom row, $p$ varied at $n=500$. The shaded configuration, $n=500$ and $p=50$, is common to both rows.}
\label{fig:sim-main-recovery}
\end{figure}

Figure~\ref{fig:sim-main-recovery} shows the TPR, FDR, and $F_1$ score of each method in the five configurations. SpaDAG attained the highest mean $F_1$ score in every configuration, with the highest TPR and the lowest FDR. ContextualDAG, the strongest comparator, had lower $F_1$ score than SpaDAG in all 150 datasets despite using the true number of edges, and the variant given the true ordering likewise had lower mean $F_1$ score in every configuration. Most of the difference lies in finding connections. SpaDAG recovered 62--96\% of the true connections in either direction, compared with 32--58\% for ContextualDAG, and chose the correct direction for 97--98\% of those it recovered, compared with 80--83\% for ContextualDAG and 53--55\% for JointGES (Table~\ref*{tab:supp-sim-decomposition}). The component-reversal move contributes to this recovery: on five baseline datasets, a sampler that performed the same computations but always rejected reversal proposals had mean $F_1$ score lower by 0.16 (Section~\ref*{sec:supp-reversal-ablation}).

Two further studies examine whether this advantage depends on orientation reversals and how recovery changes when the triangulation is built from the data. In the first, we generate every local graph as a subgraph of a single DAG, so that all locations share a topological ordering, as JointGES assumes. The edges then differ across locations only in where they are present, and we consider three levels of spatial smoothness, from edges present on contiguous regions to edges present on scattered patches (Section~\ref*{sec:supp-sim-noreversal}). SpaDAG again recovered the local graphs more accurately than every comparator, including JointGES.

The second study builds the triangulation from the observed locations, as one would in a real application to spatial transcriptomics data, using a constrained Delaunay triangulation with a resolution similar to that of the generating triangulation (Section~\ref*{subsec:supp-triangulation}). The resulting triangles cross boundaries of the true graph field, so even an oracle that knows the field cannot exceed a mean $F_1$ score of 0.894 on them. SpaDAG's $F_1$ score fell from 0.86 to 0.72 but remained substantially higher than that of every comparator, although the comparators that estimate a graph on each triangle were given the true triangulation. Coarser and finer triangulations are compared in Section~\ref*{sec:supp-mesh-resolution}.

\FloatBarrier

\section{ccRCC spatial transcriptomics analysis}
\label{sec:rcc}

We now turn to the ccRCC data introduced in Section~\ref{sec:intro-rcc} to examine how the pathway's regulatory relationships differ between the tumor and the adjacent kidney tissue. Several therapies for advanced ccRCC are directed at components of the KEGG RCC pathway: VEGF signaling, including VEGFA; HIF-2$\alpha$, which EPAS1 encodes; and MET, the receptor of HGF \citep{choueiri2020hif2vegf,choueiri2015cabozantinib,choueiri2024belzutifan}. Since these therapies act on the tumor, the relationships of primary interest are those present in tumor tissue, and the adjacent kidney tissue shows which of them are specific to the tumor. Relationships present in tumor tissue suggest which genes would respond there if the expression of a target gene were changed, and which genes act on the target. We therefore ask which genes are highly connected in the local graphs of tumor-enriched tissue and whether they are also highly connected outside it (Section~\ref{sec:rcc-hubs}), and which relationships are present only in tumor-enriched tissue (Section~\ref{sec:rcc-confined}). Given that a change in a target reaches only the genes downstream of it, we also ask whether the orientation of a gene pair changes between tissue contexts (Section~\ref{sec:rcc-orientations}).

\subsection{Preprocessing, tissue context, and inference}
\label{sec:rcc-setup}

We analyzed the 49 genes of the KEGG RCC pathway that are detected in at least 20\% of the 2,048 spots before quality control, at the 1,885 spots retained after quality control. Counts were converted to analytic Pearson residuals \citep{lause2021pearson}, which adjust for differences in library size (the total count of unique molecular identifiers, UMIs, of a spot), and the residuals were standardized for each gene, so that the causal statements below concern library-size-adjusted transcript abundance and the coefficients are in units of standard deviations. We constructed the analysis mesh as a constrained Delaunay triangulation with 85 triangles and a median of 21 spots per triangle (Sections~\ref*{subsec:supp-rcc-data} and~\ref*{subsec:supp-triangulation}).

\begin{figure}[!tbp]
\begingroup
\centering
\includegraphics[width=\linewidth]{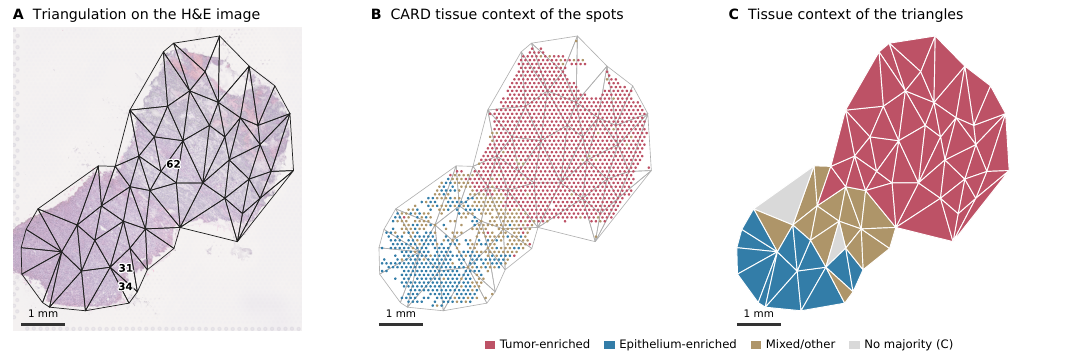}
\caption{Tissue context of the ccRCC section. (A) The triangulation over the H\&E image, with triangle numbers used in the Supplementary Material. (B) CARD tissue context of the 1,885 retained spots. (C) Tissue context of each triangle, the majority label of its spots.}
\label{fig:rcc-context}
\endgroup
\end{figure}

Each spot was labeled tumor-enriched, epithelium-enriched, or mixed/other from its CARD proportions, shown in Figure~\ref{fig:overview}A; epithelium-enriched spots are dominated by the tubular epithelium of the adjacent kidney rather than by tumor cells. Each triangle takes the majority label of its spots. Figure~\ref{fig:rcc-context}A shows the triangulation on the H\&E image, Figure~\ref{fig:rcc-context}B the labels of the spots, and Figure~\ref{fig:rcc-context}C the labels of the triangles: tumor-enriched triangles form one contiguous region, and epithelium-enriched and mixed triangles lie on the other side of the interface. Below, the tissue outside tumor-enriched tissue comprises all triangles that are not tumor-enriched, most of them epithelium-enriched or mixed/other. The definition of the labels and their sensitivity to the threshold are given in Sections~\ref*{subsec:supp-rcc-data} and~\ref*{subsec:supp-rcc-checks}.

We ran the MCMC sampler proposed in Section~\ref{sec:inference} for 50,000 iterations, discarding the first 25,000 as burn-in and retaining every fifth draw thereafter (Section~\ref*{subsec:supp-rcc-posterior}). We assessed convergence by running four independent chains, two initialized at the empty graph and two at random graphs; the results are reported in Section~\ref*{subsec:supp-rcc-chains}.
Using the PIPs from Section~\ref{sec:inference}, we selected local edges by controlling the Bayesian false discovery rate \citep[BFDR;][]{ha2021bayesian} at 0.1 (Section~\ref*{subsec:supp-edge-selection}), which yielded 78 directed relationships.

\begin{figure}[!tbp]
\begingroup
\centering
\includegraphics[width=\linewidth]{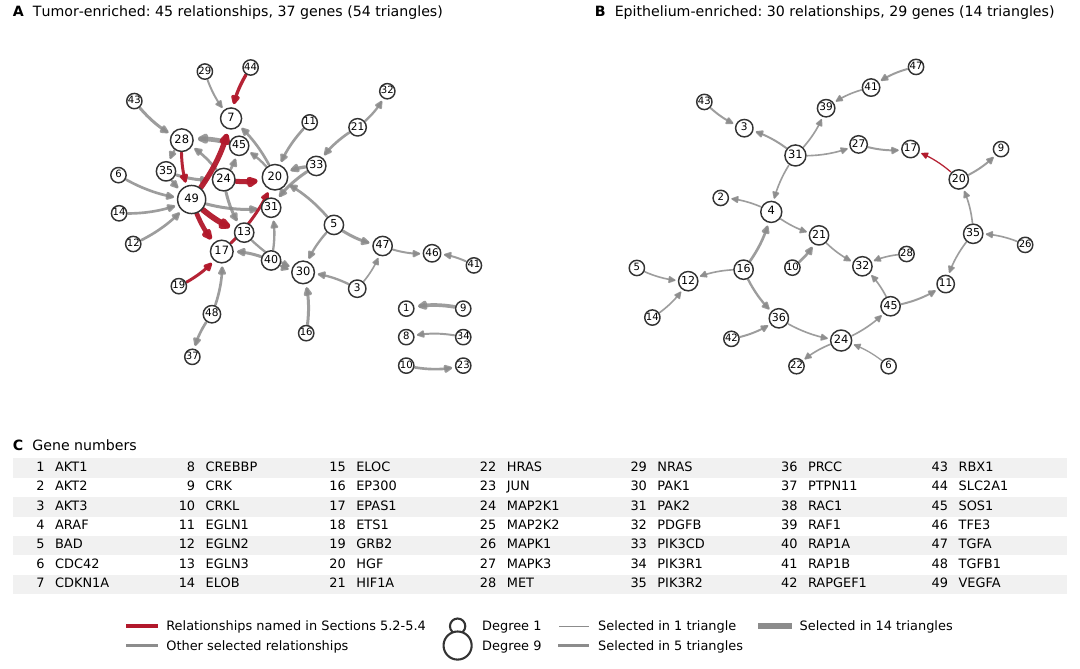}
\caption{Relationships selected at the BFDR-0.1 cutoff, by tissue context. (A, B) Relationships selected in at least one tumor-enriched triangle and in at least one epithelium-enriched triangle; node size increases with the degree of the gene, and edge width with the number of triangles in which the relationship is selected. Red edges are the relationships named in Sections~\ref{sec:rcc-hubs}--\ref{sec:rcc-orientations}. Each graph combines the relationships selected on many triangles and need not be a DAG, because acyclicity is imposed on each triangle only. The relationships selected in the mixed/other triangles are shown in Figure~\ref*{fig:supp-rcc-union-mixed}. (C) Gene numbers.}
\label{fig:rcc-union}
\endgroup
\end{figure}

Figure~\ref{fig:rcc-union}A and B summarize the selected relationships by tissue context. Panel A shows the 45 relationships selected in at least one tumor-enriched triangle, and panel B the 30 relationships selected in at least one epithelium-enriched triangle. The two graphs share only two relationships, HIF1A$\rightarrow$PDGFB and MAP2K1$\rightarrow$SOS1, so the two tissue contexts support almost entirely different directed relationships. Per triangle, the number of selected relationships is about five in tumor-enriched and in epithelium-enriched triangles alike.

\subsection{Highly connected genes in tumor-enriched tissue}
\label{sec:rcc-hubs}

Figure~\ref{fig:rcc-union}A shows that in tumor-enriched tissue, VEGFA, HGF, EPAS1, and MET are among the genes with the most selected relationships. These are the genes associated with the three therapeutic targets. About half of the relationships selected in tumor-enriched tissue involve one of the four genes, compared with about one in six outside it, the same as their share among all candidate relationships (Section~\ref*{subsec:supp-rcc-edges}). Outside tumor-enriched tissue no gene stands out in this way; the selected relationships are spread over many genes.

\begin{figure}[!tbp]
\begingroup
\centering
\includegraphics[width=\linewidth]{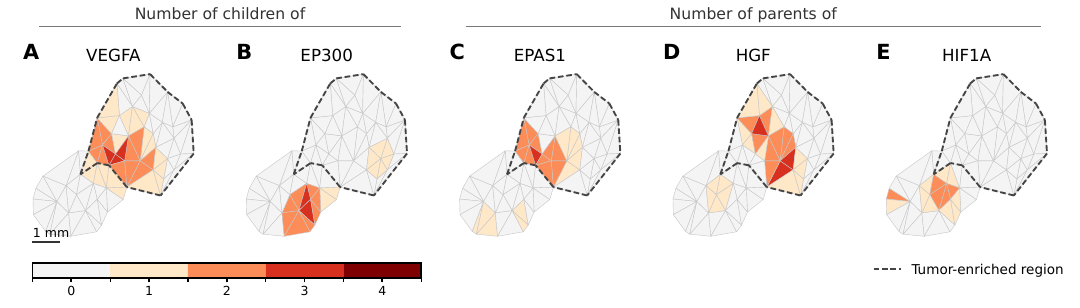}
\caption{Selected children and parents per triangle at the BFDR-0.1 cutoff. (A, B) Number of selected children of VEGFA and EP300 in each triangle. (C--E) Number of selected parents of EPAS1, HGF, and HIF1A. The dashed line encloses the tumor-enriched triangles; all tissue contexts on the same triangulation are shown in Figure~\ref{fig:rcc-context}C.}
\label{fig:rcc-hubs}
\endgroup
\end{figure}

The graphs of Figure~\ref{fig:rcc-union} combine the relationships selected in all triangles of a tissue context. A gene with many relationships in such a graph need not have many in any one triangle: it may have one child in some triangles and a different child in others. To see whether a gene is highly connected within a single local DAG, we count the selected children of each gene in each triangle, and call the gene a hub in the triangles where it has several children (Section~\ref*{subsec:supp-rcc-edges}); a change in the expression of a hub would reach several genes at once. Figure~\ref{fig:rcc-hubs}A and B show this count over the section for VEGFA and EP300, which encodes the transcriptional coactivator p300. These genes are hubs over the widest areas, VEGFA only in tumor-enriched tissue and EP300 only outside it. In tumor-enriched tissue, VEGFA acts on EPAS1, EGLN3, and CDKN1A. VEGFA and two of its children are genes of the HIF response: HIF induces VEGFA, EGLN3, and, in ccRCC, EPAS1 itself \citep{delpeso2003vhl, carroll2006hif, naas2026epas1}. The hub of tumor-enriched tissue thus connects genes of the HIF response, including those associated with two of the therapeutic targets, HIF-2$\alpha$ and VEGF signaling.

A gene can also be highly connected as a child, as a point where regulatory inputs converge. For a target, several parents mean that more than one upstream gene can change its expression. We therefore also count the selected parents of each gene in each triangle. In tumor-enriched tissue, the genes that have two or more parents over the widest areas are HGF and EPAS1. Figure~\ref{fig:rcc-hubs}C and D show the number of selected parents of EPAS1 and of HGF in each triangle. The parents of HGF include MAP2K1, a gene of the MAPK cascade, and those of EPAS1 include the adaptor gene GRB2; both genes take part in growth-factor signaling and have positive estimated effects. EPAS1 is itself a parent of HGF, a relationship that Section~\ref{sec:rcc-orientations} examines. In tumor-enriched tissue, then, several genes of the pathway act on HGF and on EPAS1 at the same place.

Neither HGF nor EPAS1 has more than one parent in any triangle outside tumor-enriched tissue. There, as Figure~\ref{fig:rcc-hubs}E shows, one of the genes with several parents over the widest areas is HIF1A, which encodes HIF-1$\alpha$ (Section~\ref*{subsec:supp-rcc-edges}). In renal tubular epithelial cells, hypoxia (low oxygen) induces HIF-1$\alpha$ accumulation, whereas VHL inactivation enables HIF-2$\alpha$ expression \citep{schietke2012hif}. HIF-2$\alpha$ supports the growth of VHL-defective ccRCC models \citep{kondo2003hif2}. In our analysis, the genes that encode the two subunits have several selected parents on opposite sides of the interface: EPAS1 in tumor-enriched tissue and HIF1A outside it.

\subsection{Relationships present only in tumor-enriched tissue}
\label{sec:rcc-confined}

Of the relationships selected only in tumor-enriched tissue, we describe two: one links two of the therapeutic targets, and the other lies downstream of HIF-2$\alpha$. The first is MET$\rightarrow$VEGFA. MET is a parent of VEGFA across a tumor-enriched region, and the relationship is not selected outside tumor-enriched tissue. It was selected in six triangles (Figure~\ref*{fig:supp-rcc-confined-pip}A), and in all six an increase of one standard deviation in MET raises VEGFA by 0.33 to 0.41 standard deviations (posterior mean coefficient averaged over each triangle). The direction is consistent with the increase in VEGF-A secretion that follows overexpression of wild-type MET in ACHN renal carcinoma cells \citep{miyata2003met}.

The second is SLC2A1$\rightarrow$CDKN1A. SLC2A1 encodes the glucose transporter GLUT1, which HIF-2$\alpha$ induces in VHL-deficient renal carcinoma cells \citep{raval2005hif}, and CDKN1A encodes the cell-cycle inhibitor p21. SpaDAG supports SLC2A1$\rightarrow$CDKN1A with posterior probability near one across a tumor-enriched region, and outside tumor-enriched tissue it supports neither direction between the two genes. The relationship was selected in seven triangles, all of them tumor-enriched (Figure~\ref*{fig:supp-rcc-confined-pip}B). The estimated direct effect is positive in all seven triangles: an increase of one standard deviation in SLC2A1 raises CDKN1A by about 0.2 standard deviations (posterior mean coefficient averaged over each triangle, 0.19 to 0.21). A dependence of both genes on HIF-2$\alpha$ would produce the opposite sign, since HIF-2$\alpha$ induces SLC2A1 but lowers p21 in renal carcinoma cells and in ccRCC tumors \citep{gordan2007hif2,gordan2008hif}; this shared dependence alone therefore cannot explain the positive coefficient. The direction from SLC2A1 to CDKN1A is consistent with the increase in p21 that follows glucose uptake in renal tubular cells \citep{kitada2014hyperglycemia} and overexpression of GLUT1 in rat kidney mesangial cells \citep{zheng2008rhein}.

\subsection{Spatially varying direction of EPAS1--HGF}
\label{sec:rcc-orientations}

\begin{figure}[!htbp]
\begingroup
\centering
\includegraphics[width=\linewidth]{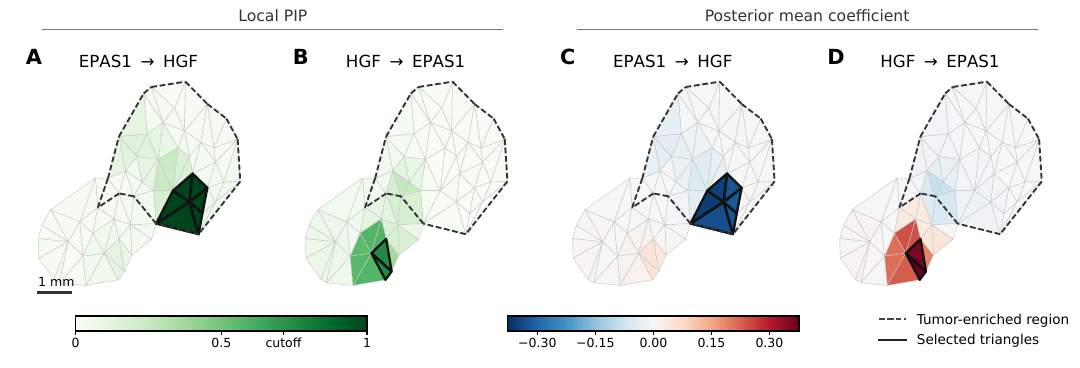}
\caption{Opposite local orientations of EPAS1--HGF. (A, B) Local PIP of each orientation over all 85 triangles. (C, D) Posterior mean coefficient of each orientation, averaged over each triangle. Solid outlines mark the triangles selected at the BFDR-0.1 cutoff, and the dashed line encloses the tumor-enriched triangles.}
\label{fig:rcc-orientation}
\endgroup
\end{figure}

The clearest spatial change in direction involves EPAS1 and HGF, the only gene pair selected in both directions. SpaDAG estimates EPAS1$\rightarrow$HGF across a tumor-enriched region and the reverse, HGF$\rightarrow$EPAS1, in one epithelium-enriched and one mixed triangle. Figure~\ref{fig:rcc-orientation}A and B show the local PIP of EPAS1$\rightarrow$HGF and of HGF$\rightarrow$EPAS1 over all 85 triangles, with the triangles selected for each direction outlined. Figure~\ref{fig:rcc-orientation}C and D show that the two estimates also differ in sign: in the six tumor-enriched triangles an increase of one standard deviation in EPAS1 lowers HGF by 0.32 to 0.36 standard deviations, and in the two triangles outside tumor-enriched tissue an increase of one standard deviation in HGF raises EPAS1 by 0.36 and 0.38 (posterior mean coefficients averaged over each triangle).

The posterior also shows how firmly each direction is supported. In the six tumor-enriched triangles the posterior probability of EPAS1$\rightarrow$HGF is near one. In the two triangles outside tumor-enriched tissue, the posterior probability is about 0.76 for HGF$\rightarrow$EPAS1. A model that imposes a single ordering on the whole tissue could represent only one of the two directions.

EPAS1 and HGF link two of the therapeutic targets, HIF-2$\alpha$ and MET signaling. Functional coupling between HGF and HIF-2$\alpha$ has been observed in a human endothelial cell line, where HIF-2$\alpha$ knockdown reduced HGF-induced IL-8 secretion \citep{makarevich2018hgf}. The estimated relationships concern their transcripts and are new hypotheses about how HIF-2$\alpha$ and MET signaling are coupled in each tissue context.

\subsection{Coordinate-permutation control}
\label{sec:rcc-permutation}

The local patterns above could reflect where expression lies in the tissue, or they could be produced by the spatial prior and the selection rule alone. To separate the two, we refitted SpaDAG after randomly permuting the spot coordinates once. This keeps every expression profile but assigns the profiles to locations at random, so that every region draws its profiles from the same pooled set of spots (Section~\ref*{subsec:supp-rcc-checks}).

Without the spatial arrangement, the fitted graphs lost their local structure. Of the 78 relationships selected with the observed coordinates, 68, including both directions of EPAS1--HGF, were selected on no triangle after permutation (Section~\ref*{subsec:supp-rcc-checks} and Table~\ref*{tab:supp-rcc-stability}). The variation of posterior probabilities across triangles also largely disappeared, and a relationship whose two genes covary across the whole section, such as VEGFA$\rightarrow$EGLN3, became supported almost everywhere (Figure~\ref*{fig:supp-rcc-permutation}). The localized patterns of Sections~\ref{sec:rcc-hubs}--\ref{sec:rcc-orientations} therefore depend on the spatial arrangement of expression.

\section{Discussion}
\label{sec:discussion}

In this article, we propose SpaDAG, which \rm infers a causal DAG at each location of a tissue section, with posterior probabilities for the presence and the direction of each edge. Applying SpaDAG to a ccRCC section, we found that \rm tumor-enriched and epithelium-enriched tissue supported almost entirely different directed relationships, and that \rm the estimated direction between EPAS1 and HGF differed between a tumor-enriched region and two triangles outside it. These relationships are hypotheses at the transcript level, which experiments at the protein level can test within each tissue context.

Our interpretation \rm rests on location-wise causal sufficiency (Section~\ref{subsec:spatial-sem-fe}). Unmeasured regulators, such as those acting at the protein level, can be common causes of the modeled genes; for example, VEGFA$\rightarrow$EGLN3 may reflect the regulation of both genes by HIF. In spot-based data such as those analyzed here, the mixture of cell types within a spot can act in the same way. Our identifiability result assumes \rm error variances that are constant over space (Section~\ref{subsec:identifiability}). In such data, sequencing depth and cell composition differ between spots, so noise levels can vary within a tissue.

SpaDAG fixes the triangulation \rm before sampling; updating it within the sampler would let the boundaries of the triangles follow the changes in the local graphs. The sampler can also be extended with moves that add or remove an edge over a connected region. Although motivated by spatial transcriptomics, SpaDAG can be applied to other spatially resolved multivariate measurements, such as spatial proteomics, when the assumptions of the spatial SEM are appropriate.

\FloatBarrier
\setlength{\bibsep}{0pt}

\end{document}